

\magnification=\magstephalf
\hsize=13.0 true cm
\vsize=19 true cm
\hoffset=1.50 true cm
\voffset=2.0 true cm

\abovedisplayskip=12pt plus 3pt minus 3pt
\belowdisplayskip=12pt plus 3pt minus 3pt
\parindent=2em


\font\sixrm=cmr6
\font\eightrm=cmr8
\font\ninerm=cmr9

\font\sixi=cmmi6
\font\eighti=cmmi8
\font\ninei=cmmi9

\font\sixsy=cmsy6
\font\eightsy=cmsy8
\font\ninesy=cmsy9

\font\sixbf=cmbx6
\font\eightbf=cmbx8
\font\ninebf=cmbx9

\font\eightit=cmti8
\font\nineit=cmti9

\font\eightsl=cmsl8
\font\ninesl=cmsl9

\font\sixss=cmss8 at 8 true pt
\font\sevenss=cmss9 at 9 true pt
\font\eightss=cmss8
\font\niness=cmss9
\font\tenss=cmss10

\font\bigrm=cmr10 at 12 true pt
\font\bigbf=cmbx10 at 12 true pt

\catcode`@=11
\newfam\ssfam

\def\tenpoint{\def\rm{\fam0\tenrm}%
    \textfont0=\tenrm \scriptfont0=\sevenrm \scriptscriptfont0=\fiverm
    \textfont1=\teni  \scriptfont1=\seveni  \scriptscriptfont1=\fivei
    \textfont2=\tensy \scriptfont2=\sevensy \scriptscriptfont2=\fivesy
    \textfont3=\tenex \scriptfont3=\tenex   \scriptscriptfont3=\tenex
    \textfont\itfam=\tenit                  \def\it{\fam\itfam\tenit}%
    \textfont\slfam=\tensl                  \def\sl{\fam\slfam\tensl}%
    \textfont\bffam=\tenbf \scriptfont\bffam=\sevenbf
    \scriptscriptfont\bffam=\fivebf
                                            \def\bf{\fam\bffam\tenbf}%
    \textfont\ssfam=\tenss \scriptfont\ssfam=\sevenss
    \scriptscriptfont\ssfam=\sevenss
                                            \def\ss{\fam\ssfam\tenss}%
    \normalbaselineskip=13pt
    \setbox\strutbox=\hbox{\vrule height8.5pt depth3.5pt width0pt}%
    \let\big=\tenbig
    \normalbaselines\rm}

\def\ninepoint{\def\rm{\fam0\ninerm}%
    \textfont0=\ninerm      \scriptfont0=\sixrm
                            \scriptscriptfont0=\fiverm
    \textfont1=\ninei       \scriptfont1=\sixi
                            \scriptscriptfont1=\fivei
    \textfont2=\ninesy      \scriptfont2=\sixsy
                            \scriptscriptfont2=\fivesy
    \textfont3=\tenex       \scriptfont3=\tenex
                            \scriptscriptfont3=\tenex
    \textfont\itfam=\nineit \def\it{\fam\itfam\nineit}%
    \textfont\slfam=\ninesl \def\sl{\fam\slfam\ninesl}%
    \textfont\bffam=\ninebf \scriptfont\bffam=\sixbf
                            \scriptscriptfont\bffam=\fivebf
                            \def\bf{\fam\bffam\ninebf}%
    \textfont\ssfam=\niness \scriptfont\ssfam=\sixss
                            \scriptscriptfont\ssfam=\sixss
                            \def\ss{\fam\ssfam\niness}%
    \normalbaselineskip=12pt
    \setbox\strutbox=\hbox{\vrule height8.0pt depth3.0pt width0pt}%
    \let\big=\ninebig
    \normalbaselines\rm}

\def\eightpoint{\def\rm{\fam0\eightrm}%
    \textfont0=\eightrm      \scriptfont0=\sixrm
                             \scriptscriptfont0=\fiverm
    \textfont1=\eighti       \scriptfont1=\sixi
                             \scriptscriptfont1=\fivei
    \textfont2=\eightsy      \scriptfont2=\sixsy
                             \scriptscriptfont2=\fivesy
    \textfont3=\tenex        \scriptfont3=\tenex
                             \scriptscriptfont3=\tenex
    \textfont\itfam=\eightit \def\it{\fam\itfam\eightit}%
    \textfont\slfam=\eightsl \def\sl{\fam\slfam\eightsl}%
    \textfont\bffam=\eightbf \scriptfont\bffam=\sixbf
                             \scriptscriptfont\bffam=\fivebf
                             \def\bf{\fam\bffam\eightbf}%
    \textfont\ssfam=\eightss \scriptfont\ssfam=\sixss
                             \scriptscriptfont\ssfam=\sixss
                             \def\ss{\fam\ssfam\eightss}%
    \normalbaselineskip=10pt
    \setbox\strutbox=\hbox{\vrule height7.0pt depth2.0pt width0pt}%
    \let\big=\eightbig
    \normalbaselines\rm}

\def\tenbig#1{{\hbox{$\left#1\vbox to8.5pt{}\right.\n@space$}}}
\def\ninebig#1{{\hbox{$\textfont0=\tenrm\textfont2=\tensy
                       \left#1\vbox to7.25pt{}\right.\n@space$}}}
\def\eightbig#1{{\hbox{$\textfont0=\ninerm\textfont2=\ninesy
                       \left#1\vbox to6.5pt{}\right.\n@space$}}}

\font\sectionfont=cmbx10
\font\subsectionfont=cmti10

\def\figurecaptionfont{\ninepoint}
\def\tablecaptionfont{\ninepoint}
\def\footnotefont{\eightpoint}


\newcount\equationno
\newcount\bibitemno
\newcount\figureno
\newcount\tableno

\equationno=0
\bibitemno=0
\figureno=0
\tableno=0
\advance\pageno by -1


\footline={\ifnum\pageno=0{\hfil}\else
{\hss\rm\the\pageno\hss}\fi}


\def\section #1. #2 \par
{\vskip0pt plus .20\vsize\penalty-150 \vskip0pt plus-.20\vsize
\vskip 1.6 true cm plus 0.2 true cm minus 0.2 true cm
\global\def\equationlabel{#1}
\global\equationno=0
\centerline{\sectionfont #1. #2}\par
\immediate\write\terminal{Section #1. #2}
\vskip 0.7 true cm plus 0.1 true cm minus 0.1 true cm}


\def\subsection #1 \par
{\vskip0pt plus .15\vsize\penalty-50 \vskip0pt plus-.15\vsize
\vskip2.5ex plus 0.1ex minus 0.1ex
\leftline{\subsectionfont #1}\par
\immediate\write\terminal{Subsection #1}
\vskip1.0ex plus 0.1ex minus 0.1ex
\noindent}


\def\appendix #1 \par
{\vskip0pt plus .20\vsize\penalty-150 \vskip0pt plus-.20\vsize
\vskip 1.6 true cm plus 0.2 true cm minus 0.2 true cm
\global\def\equationlabel{\hbox{\rm#1}}
\global\equationno=0
\centerline{\sectionfont Appendix #1}\par
\immediate\write\terminal{Appendix #1}
\vskip 0.7 true cm plus 0.1 true cm minus 0.1 true cm}


\def\enum{\global\advance\equationno by 1
(\equationlabel.\the\equationno)}


\def\ifundefined#1{\expandafter\ifx\csname#1\endcsname\relax}

\def\ref#1{\ifundefined{#1}?\immediate\write\terminal{unknown reference
on page \the\pageno}\else\csname#1\endcsname\fi}

\newwrite\terminal
\newwrite\bibitemlist

\def\bibitem#1#2\par{\global\advance\bibitemno by 1
\immediate\write\bibitemlist{\string\def
\expandafter\string\csname#1\endcsname
{\the\bibitemno}}
\item{[\the\bibitemno]}#2\par}

\def\beginbibliography{
\vskip0pt plus .20\vsize\penalty-150 \vskip0pt plus-.20\vsize
\vskip 1.6 true cm plus 0.2 true cm minus 0.2 true cm
\centerline{\sectionfont References}\par
\immediate\write\terminal{References}
\immediate\openout\bibitemlist=biblist.aux
\frenchspacing
\vskip 0.7 true cm plus 0.1 true cm minus 0.1 true cm}

\def\endbibliography{
\immediate\closeout\bibitemlist
\nonfrenchspacing}


\def\figurecaption#1{\global\advance\figureno by 1
\narrower\figurecaptionfont
Fig.~\the\figureno. #1}

\def\tablecaption#1{\global\advance\tableno by 1
\vbox to 0.5 true cm { }
\centerline{\tablecaptionfont%
Table~\the\tableno. #1}
\vskip-0.4 true cm}

\tenpoint

\immediate\openin\bibitemlist=biblist.aux
\ifeof\bibitemlist\immediate\closein\bibitemlist
\else\immediate\closein\bibitemlist
\input biblist.aux \fi


\def\blackboardrrm{\mathchoice
{\rm I\kern-0.21 em{R}}{\rm I\kern-0.21 em{R}}
{\rm I\kern-0.19 em{R}}{\rm I\kern-0.19 em{R}}}

\def\blackboardzrm{\mathchoice
{\rm Z\kern-0.32 em{Z}}{\rm Z\kern-0.32 em{Z}}
{\rm Z\kern-0.28 em{Z}}{\rm Z\kern-0.28 em{Z}}}

\def\blackboardh{\mathchoice
{\ss I\kern-0.14 em{H}}{\ss I\kern-0.14 em{H}}
{\ss I\kern-0.11 em{H}}{\ss I\kern-0.11 em{H}}}

\def\blackboardp{\mathchoice
{\ss I\kern-0.14 em{P}}{\ss I\kern-0.14 em{P}}
{\ss I\kern-0.11 em{P}}{\ss I\kern-0.11 em{P}}}

\def\blackboardt{\mathchoice
{\ss T\kern-0.52 em{T}}{\ss T\kern-0.52 em{T}}
{\ss T\kern-0.40 em{T}}{\ss T\kern-0.40 em{T}}}

\def\thicktablerule{\hrule height1pt}
\def\thintablerule{\hrule height0.4pt}



\def\rme{{\rm e}}
\def\rmO{{\rm O}}




\def\proof{\noindent{\sl Proof:}\kern0.6em}

\def\frac#1#2{\hbox{$#1\over#2$}}
\def\dual{\mathstrut^*\kern-0.1em}

\def\lvec#1{\setbox0=\hbox{$#1$}
    \setbox1=\hbox{$\scriptstyle\leftarrow$}
    #1\kern-\wd0\smash{
    \raise\ht0\hbox{$\raise1pt\hbox{$\scriptstyle\leftarrow$}$}}
    \kern-\wd1\kern\wd0}
\def\rvec#1{\setbox0=\hbox{$#1$}
    \setbox1=\hbox{$\scriptstyle\rightarrow$}
    #1\kern-\wd0\smash{
    \raise\ht0\hbox{$\raise1pt\hbox{$\scriptstyle\rightarrow$}$}}
    \kern-\wd1\kern\wd0}


\def\nabstar#1{\nabla\kern-0.5pt\smash{\raise 4.5pt\hbox{$\ast$}}
               \kern-4.5pt_{#1}}
\def\drv#1{{\partial_{#1}}}
\def\drvstar#1{\partial\kern-0.5pt\smash{\raise 4.5pt\hbox{$\ast$}}
               \kern-5.0pt_{#1}}


\def\momp#1#2{
    \setbox0=\hbox{${#1}$}\setbox1=\hbox{${#1}_{#2}$}
    {#1}_{#2}\kern-\wd1\kern\wd0
    \smash{\raise4.5pt\hbox{$\scriptscriptstyle +$}}}
\def\momm#1#2{
    \setbox0=\hbox{${#1}$}\setbox1=\hbox{${#1}_{#2}$}
    {#1}_{#2}\kern-\wd1\kern\wd0
    \smash{\raise4.5pt\hbox{$\scriptscriptstyle -$}}}
\def\mompm#1#2{
    \setbox0=\hbox{${#1}$}\setbox1=\hbox{${#1}_{#2}$}
    {#1}_{#2}\kern-\wd1\kern\wd0
    \smash{\raise4.5pt\hbox{$\scriptscriptstyle \pm$}}}
\def\smomp#1#2{
    \setbox0=\hbox{${#1}$}\setbox1=\hbox{${#1}_{#2}$}
    {#1}_{#2}\kern-\wd1\kern\wd0
    \smash{\raise3pt\hbox{$\scriptscriptstyle +$}}}
\def\smomm#1#2{
    \setbox0=\hbox{${#1}$}\setbox1=\hbox{${#1}_{#2}$}
    {#1}_{#2}\kern-\wd1\kern\wd0
    \smash{\raise3pt\hbox{$\scriptscriptstyle -$}}}
\def\smompm#1#2{
    \setbox0=\hbox{${#1}$}\setbox1=\hbox{${#1}_{#2}$}
    {#1}_{#2}\kern-\wd1\kern\wd0
    \smash{\raise3pt\hbox{$\scriptscriptstyle \pm$}}}

\def\si{\kern1pt{\rm si}}
\def\co{\kern1pt{\rm co}}




\def\Nf{N_{\rm f}}

\def\psiclass{\psi_{\rm cl}}

\def\rhoprime{\rho\kern1pt'}
\def\rhobar{\bar{\rho}}
\def\rhobarprime{\rhobar\kern1pt'}
\def\rhobartilde{\kern2pt\tilde{\kern-2pt\rhobar}}
\def\rhobartildeprime{\kern2pt\tilde{\kern-2pt\rhobar}\kern1pt'}

\def\zetabar{\bar{\zeta}}
\def\zetaprime{\zeta\kern1pt'}
\def\zetabarprime{\zetabar\kern1pt'}
\def\zetar{\zeta_{\raise-1pt\hbox{\sixrm R}}}
\def\zetabarr{\zetabar_{\raise-1pt\hbox{\sixrm R}}}

\def\phiimpr{\phi_{\kern0.5pt\hbox{\sixrm I}}}


\def\dirac#1{\gamma_{#1}}
\def\diracstar#1#2{
    \setbox0=\hbox{$\gamma$}\setbox1=\hbox{$\gamma_{#1}$}
    \gamma_{#1}\kern-\wd1\kern\wd0
    \smash{\raise4.5pt\hbox{$\scriptstyle#2$}}}


\def\ba{b_{\rm A}}
\def\bp{b_{\rm P}}
\def\bv{b_{\rm V}}
\def\bt{b_{\rm T}}

\def\bg{b_{\rm g}}
\def\bm{b_{\rm m}}
\def\bzeta{b_{\zeta}}

\def\ca{c_{\rm A}}
\def\cv{c_{\rm V}}
\def\Ca{\ca}
\def\Cv{\cv}
\def\csw{c_{\rm sw}}

\def\cst{c_{\rm s,t}}

\def\ctildet{\tilde{c}_{\rm t}}
\def\ctildest{\tilde{c}_{\rm s,t}}


\def\fa{f_{\rm A}}

\def\fp{f_{\rm P}}

\def\kv{k_{\rm V}}
\def\kt{k_{\rm T}}
\def\f1{f_1}

\def\ha{h_{\rm A}}
\def\hda{h_{\rm dA}}
\def\hp{h_{\rm P}}
\def\hv{h_{\rm V}}
\def\h1{h_1}


\def\SUthree{{\rm SU(3)}}
\def\SUn{{\rm SU}(N)}
\def\tr{\,\hbox{tr}\,}

\def\CF{C_{\rm F}}
\def\cf{\CF}


\def\opprime#1{\setbox0=\hbox{${\cal O}$}\setbox1=\hbox{${\cal O}_{\rm #1}$}
    {\cal O}_{\rm #1}\kern-\wd1\kern\wd0
    \smash{\raise4.5pt\hbox{\kern1pt$\scriptstyle\prime$}}\kern1pt}

\def\ophatprime#1{\setbox0=\hbox{$\widehat{\cal O}$}
    \setbox1=\hbox{$\widehat{\cal O}_{\rm #1}$}
    \widehat{\cal O}_{\rm #1}\kern-\wd1\kern\wd0
    \smash{\raise4.5pt\hbox{\kern1pt$\scriptstyle\prime$}}\kern1pt}

\def\bopprime#1{\setbox0=\hbox{${\cal O}$}\setbox1=\hbox{${\cal O}_{\rm #1}$}
    {\cal L}_{\rm #1}\kern-\wd1\kern\wd0
    \smash{\raise4.5pt\hbox{\kern1pt$\scriptstyle\prime$}}\kern1pt}

\def\blagprime#1{\setbox0=\hbox{${\cal B}$}\setbox1=\hbox{${\cal B}_{#1}$}
    {\cal B}_{#1}\kern-\wd1\kern\wd0
    \smash{\raise5.2pt\hbox{\kern1pt$\scriptstyle\prime$}}\kern1pt}


\def\gr{g_{{\hbox{\sixrm R}}}}
\def\gR{\gr}

\def\gp{g_{\rm P}}

\def\mq{m_{\rm q}}

\def\mr{m_{{\hbox{\sixrm R}}}}
\def\mc{m_{\rm c}}

\def\za{Z_{\rm A}}
\def\zv{Z_{\rm V}}
\def\zp{Z_{\rm P}}

\def\zm{Z_{\rm m}}

\def\zzeta{Z_{\zeta}}

\def\Zv{\zv}
\def\Zp{\zp}

\def\Zz{\zzeta}

\def\gtilde{\tilde{g}_0}

\def\msbar{{\rm \overline{MS\kern-0.05em}\kern0.05em}}

\def\SF{\rm SF}

\input epsf
\rightline{MPI-PhT/97-21}
\rightline{FSU-SCRI-97-24}
\vskip 1.0 true cm minus 0.3 true cm
\centerline
{\bigbf Further results on O(a) improved lattice QCD}
\vskip2ex
\centerline
{\bigbf to one-loop order of perturbation theory}
\vskip 1.5 true cm
\centerline{\bigrm Stefan Sint}
\vskip1ex
\centerline{SCRI, The Florida State University,}
\centerline{Tallahassee 32306--4052, Florida, U.S.A.}
\vskip 0.8 true cm
\centerline{\bigrm Peter Weisz}
\vskip1ex
\centerline{Max-Planck-Institut f\"ur Physik}
\centerline{F\"ohringer Ring 6, D-80805 M\"unchen, Germany}
\vskip 1.5 true cm
\centerline{\bf Abstract}
\vskip 1.5ex
We present results at one-loop order of perturbation theory for
various improvement coefficients in on-shell
O($a$) improved lattice QCD. In particular we determine 
the additive counterterm required for on-shell
improvement of the isovector vector current.
Employing a general mass-independent renormalization scheme
we also obtain the  coefficients of the O($a$) counterterms
which are proportional to the quark mass in the improved
isovector pseudo-scalar, axial vector and vector operators.
In the latter case a comparison with a recent non-perturbative
study is made.
\vfill
\centerline{April 1997}
\eject

\section 1. Introduction

This paper belongs to a series of publications on 
chiral symmetry and O($a$) improvement in lattice
QCD with Wilson quarks~[\ref{letter}--\ref{paperIV}].
Here we report on the computation of various
improvement coefficients to one-loop order of
perturbation theory.
The underlying theoretical framework is based on the Schr\"odinger
functional~(SF)~[\ref{alphaI},\ref{StefanI}] and has been presented 
in detail in the papers referred to above. In particular, a general overview 
can be obtained from refs.~[\ref{letter},\ref{latconf}].

The present paper is organized as follows:
In section~2 we recall various definitions
limiting ourselves to those which are essential for the understanding
of this paper. These include in particular the definitions of
the on-shell O($a$) improved isovector pseudo-scalar, 
axial vector and vector operators, and of the renormalized 
parameters and fields in a mass-independent 
renormalization scheme. We then summarize our 
main results which are contained in table~1 and eq.~(2.16).

In section~3 we present a few details of the one-loop 
calculation and discuss the renormalization procedure for 
the general case of non-vanishing renormalized quark mass.
Section~4 contains the determination of the improved vector
current and of various improvement coefficients which
arise in the case of non-zero quark mass.
We end with a few concluding remarks~(section~5).
Finally, two appendices have been included which
provide analytic expressions for the tree-level
amplitudes and counterterms at zero and non-zero
quark mass respectively.

\section 2. Overview and results 

A general introduction to on-shell O($a$) improvement 
of lattice QCD with Wilson quarks can be found in ref.~[\ref{paperI}].
In particular there it has been discussed how
the renormalization procedure can be carried out in a way consistent
with O($a$) improvement. In the following we adopt 
the conventions and notations as introduced in this reference. 

\subsection 2.1 Definitions

On-shell O($a$) improvement requires the introduction of
O($a$) counterterms for both the lattice action and the composite
fields of interest. 
The improved lattice action can be obtained by including
the  Sheikholeslami-Wohlert interaction term 
(with coefficient $\csw$)~[\ref{SW}].
Concerning the composite fields we will restrict attention
to a few gauge invariant combinations which are bilinear
in the quark fields. Assuming $\Nf\ge 2$ mass degenerate
quark flavours we consider the local isovector
fields~\footnote{$^\dagger$}{\footnotefont
  The Dirac gamma matrix conventions are as in
  appendix A of ref.~[\ref{paperI}]},
$$
  \eqalignno{
  V^a_{\mu}(x)=&\bar{\psi}(x) \gamma_{\mu}\frac12\tau^a\psi (x),
  &\enum\cr
  \noalign{\vskip2ex}
  A^a_{\mu}(x)=&\bar{\psi}(x) \gamma_{\mu}\gamma_5\frac12\tau^a\psi (x),
  &\enum\cr
  \noalign{\vskip2ex}
  P^a (x)=&\bar{\psi}(x) \gamma_5\frac12\tau^a\psi (x),
  &\enum\cr
  \noalign{\vskip2ex}
  T^a_{\mu\nu}(x)=&i\bar{\psi}(x) \sigma_{\mu\nu}\frac12\tau^a\psi (x).
  &\enum}
$$
Using these definitions and following ref.~[\ref{paperI}] the
improved isovector vector and axial vector
currents may be parametrized as follows,  
$$
\eqalignno{
(V_{\rm I})^a_{\mu}=&V^a_{\mu}+\cv(g_0^2) a\frac12
(\partial^*_{\nu}+\partial_{\nu})T^a_{\mu\nu},
&\enum\cr
\noalign{\vskip2ex}
(A_{\rm I})^a_{\mu}=&A^a_{\mu}+\ca(g_0^2) a\frac12
(\partial^*_{\mu}+\partial_{\mu})P^a.
&\enum}
$$

Throughout the paper we will use a mass-independent renormalization
scheme which is compatible with O($a$) improvement.
In such a scheme the renormalized coupling and quark mass
are given by~[\ref{paperI}]
$$
  \eqalignno{
  \gr^2=&\tilde{g}_0^2Z_{\rm g}(\tilde{g}_0^2,a\mu),
  &\enum\cr
  \noalign{\vskip2ex}
  m_{\rm R}=&\tilde{m}_{\rm q}\zm(\tilde{g}_0^2,a\mu),
  &\enum}
$$
where $\mu$ is the renormalization scale. 
The parameters $\tilde{g}_0$
and $\tilde{m}_{\rm q}$ are defined as follows
$$
  \eqalignno{
  \tilde{g}_0^2=&g_0^2 \bigl[ 1+\bg(g_0^2)a\mq \bigr],
  &\enum\cr
  \noalign{\vskip2ex}
  \tilde{m}_{\rm q}=&\mq\bigl[ 1+\bm(g_0^2)a\mq \bigr],\qquad \mq=m_0-\mc,
  &\enum}
$$
where $\mc$ is the critical bare quark mass.

With these definitions 
the renormalized improved currents 
and pseudo-scalar density take the form~[\ref{paperI}],
$$
  \eqalignno{
  (V_{\rm R})^a_{\mu}=&\zv(\tilde{g}_0^2,a\mu)
            \bigl[1+\bv(g_0^2) a\mq\bigr](V_{\rm I})^a_{\mu},
  &\enum\cr
  \noalign{\vskip2ex}
  (A_{\rm R})^a_{\mu}=&\za(\tilde{g}_0^2,a\mu)
	    \bigl[1+\ba(g_0^2) a\mq\bigr](A_{\rm I})^a_{\mu},
  &\enum\cr
  \noalign{\vskip2ex}
  (P_{\rm R})^a=&\zp(\tilde{g}_0^2,a\mu)\bigl[1+\bp(g_0^2) a\mq\bigr]P^a.
  &\enum}
$$

In the massless theory on-shell correlation functions involving only 
these fields are O($a$) improved if the improvement coefficients 
$\csw$, $\ca$ and $\cv$ are properly chosen as functions
of the bare coupling $g_0$. In the presence of 
massive quarks one also needs to know the various
$b$-coefficients introduced above.

To compute these coefficients 
we will consider correlation functions 
derived from the Schr\"odinger functional which involve the 
above operators and also the boundary quark fields $\zeta,\zetabar$ 
and $\zeta^{\prime},\zetabar^{\prime}$~[\ref{alphaI}]. 
O($a$) improved correlation functions
are then obtained with the renormalized fields 
$\zeta_{\rm R},\zetabar_{\rm R}$ and
$\zeta_{\rm R}^{\prime},\zetabar_{\rm R}^{\prime}$,
which are all related to the bare fields
by the  same renormalization factor, e.g.
$$
  \zeta_{\rm R}=\zzeta(\tilde{g}_0^2,a\mu)
	 	  \bigl[1+\bzeta(g_0^2) a\mq\bigr]\zeta. 
\eqno\enum
$$

For completeness we recall that Schr\"odinger functional boundary
conditions lead to additional cutoff effects which can
be cancelled by counterterms that are localized at
the boundaries. These boundary counterterms
can be chosen such that the associated improvement coefficients
$\cst$ and $\ctildest$
appear as weight factors for certain terms of the lattice action
close to the boundaries~[\ref{alphaI},\ref{paperI}].

\subsection 2.2 Results

The computation of $\csw$ to one-loop order of perturbation theory
was first performed in ref.~[\ref{Wohlert}] and the numerical result
was confirmed to three significant
digits in ref.~[\ref{paperII}]. For the purpose of our one-loop
calculation it is sufficient to know that $\csw=1$ at tree-level
of perturbation theory. 

The improvement coefficients $\ca$ and
$\cv$ are given by
$$ 
  \eqalignno{
  \ca&=\ca^{(1)}g_0^2+\rmO(g_0^4),\qquad \ca^{(1)}=-0.005680(2)\times\CF,
  &\enum\cr
  \noalign{\vskip1ex}
  \cv&=\cv^{(1)}g_0^2+\rmO(g_0^4),\qquad \cv^{(1)}=-0.01225(1)\times\CF,
  &\enum\cr}
$$
where $\CF=(N^2-1)/2N$ for gauge group $\SUn$.
The coefficient $\ca^{(1)}$ has first been obtained in ref.~[\ref{paperII}]
and the computation of $\cv^{(1)}$ will be presented in section~4.
The extended numerical data produced in the course of this
calculation also led to the improved estimate 
for $\ca^{(1)}$ as given in eq.~(2.15).


\topinsert
\newdimen\digitwidth
\setbox0=\hbox{\rm 0}
\digitwidth=\wd0
\catcode`@=\active
\def@{\kern\digitwidth}
\tablecaption{
Tree level and one-loop improvement coefficients $b$  
}
\vskip1ex
$$\vbox{\settabs\+xx&xxxxx&xxxx&xxxxxx
                          &xxxx&xxxxxxxxxxxxxxxxxx&x&\cr
\thicktablerule
\vskip1ex
                \+& \hfill ${\rm X}$ \hfill
                 && \hfill $b_{\rm X}^{(0)}$ \hfill
                 && \hfill $b_{\rm X}^{(1)}$ \hfill
                 &&  \cr
\vskip 0.5ex
\thintablerule
\vskip 1.5ex
  \+& \hfill ${\rm g}$ \hfill
  && \hfill $0$ \hfill
  && \hfill $0.012000(2)\times\Nf$\hfill
                   &&\cr
\vskip 1ex
  \+& \hfill ${\rm m}$ \hfill
  && \hfill $-\frac12$ \hfill
  && \hfill $ -0.07217(2)\times\CF$ \hfill
                   &&\cr
\vskip 1ex
  \+& \hfill $\zeta$ \hfill
  &&  \hfill $-\frac12$ \hfill
  && \hfill$ -0.06738(4)\times\CF$ \hfill
                   &&\cr
\vskip 1ex
  \+& \hfill ${\rm A}$ \hfill
  && \hfill $1$ \hfill
  && \hfill $\phantom{-}0.11414(4)\times\CF$ \hfill
                   &&\cr
\vskip 1ex
  \+& \hfill ${\rm V}$ \hfill                             
  && \hfill $1$ \hfill
  && \hfill$\phantom{-}0.11492(4)\times\CF$ \hfill                 
                   &&\cr
\vskip 1ex
  \+& \hfill ${\rm P}$ \hfill                             
  && \hfill $1$ \hfill
  && \hfill$\phantom{-}0.11484(2)\times\CF$ \hfill                             
                   &&\cr
\vskip 1ex
  \+& \hfill ${\rm T}$ \hfill                             
  && \hfill $1$ \hfill
  &&                             
                   &&\cr
\vskip 1ex
\thicktablerule
\vskip 1ex
}$$
\endinsert

For the various $b$-coefficients, whose perturbation expansion reads
$$
  b=b^{(0)}+b^{(1)}g_0^2+\rmO(g_0^4),
  \eqno\enum
$$ 
we have collected the known results in table~1. 
The computation of $\bg^{(1)}$ was performed in
ref.~[\ref{StefanRainer}] and the calculation 
of the remaining coefficients is the topic of section~4.
It is remarkable that the coefficients $\ba^{(1)},\bv^{(1)}$
and $\bp^{(1)}$ are so close numerically. To our knowledge there is, 
however, no obvious theoretical explanation of this fact.
Note also that all coefficients appear to be reasonably small,
i.e.~of $\rmO(1)$ when expanded in powers of $g_0^2/4\pi$
for gauge group $\SUthree$.

\topinsert
\vbox{
\vskip0.0true cm

\centerline{
\epsfxsize=11 true cm
\epsfbox{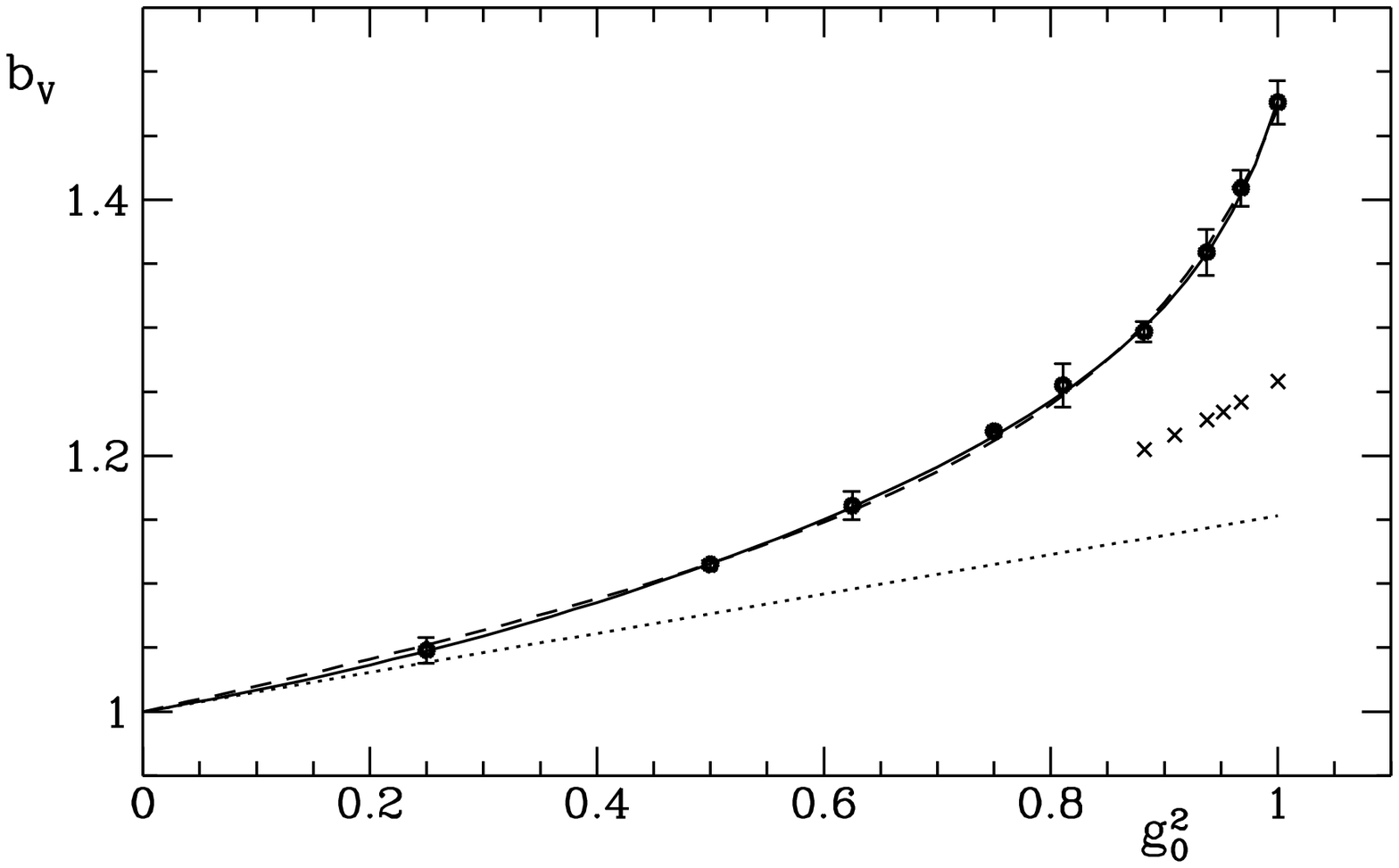}}

\vskip-1.5true cm
\figurecaption{The improvement coefficient $\bv$ as a function
of the bare coupling for gauge group $\SUthree$. The data points
and the fit function corresponding to the dashed curve 
are taken from ref.~[\ref{paperIV}]. The solid curve is the new
fit function~(2.18), the dotted line is the perturbative result
and the crosses denote the perturbative result when the
coupling $\gp$ is used instead of $g_0$.}
}
\endinsert

In the case of $\bv$ we can now make a comparison with the 
non-perturbative result for $\Nf=0$ and 
gauge group $\SUthree$ which was obtained in ref.~[\ref{paperIV}].
An updated fit of the data, which reduces to the correct
one-loop expression for small $g_0$, is given 
by~\footnote{$^\dagger$}{\footnotefont 
  We thank Hartmut Wittig for providing this fit and fig.~1.}
$$
  \bv (g_0^2)={1 - 0.7613g_0^2 + 0.0012g_0^4 - 0.1136g_0^6
  \over 1 - 0.9145g_0^2}, \qquad 0\leq g_0\leq 1. 
  \eqno\enum
$$
The data together with the fit functions are displayed in fig.~1. 
Note that both curves are indistinguishable for all practical purposes.

Using the one-loop result for $\bv$ we may also compare 
the data with so-called boosted perturbation theory. To this order
this simply amounts to replacing the bare coupling $g_0$ by
Parisi's coupling $\gp^{}$ [\ref{Parisi}], defined by
$$
  \gp^2=g_0^2/P,
  \eqno\enum
$$
where $P$ denotes the expectation value of the plaquette
in infinite volume. The corresponding values of $\bv$
are plotted as crosses in fig.~1. Although the use of $\gp^{}$
moves the perturbative result towards the non-perturbative
curve, there remains a significant discrepancy at the larger 
values of the bare coupling. 

\section 3. The one-loop calculation

We define the relevant correlation functions and provide 
some details of their evaluation to one-loop order of perturbation
theory.
This is followed by a discussion of the renormalization procedure
for the general case of non-vanishing renormalized quark
mass. In the following the reader is assumed 
to be familiar with ref.~[\ref{paperII}].
In particular the notation and the general set-up of
perturbation theory on a finite lattice with 
Schr\"odinger functional boundary conditions will be taken
over from this reference.

\subsection 3.1 Definition of the correlation functions

We consider the following correlation functions
which were introduced in refs.~[\ref{letter},\ref{paperI}], 
$$
\eqalignno{
  \fa(x_0)&=-a^6\sum_{\bf y,z}
  \frac{1}{3}\langle A_0^a(x)\,
  \zetabar({\bf y})\dirac{5}\frac{1}{2}\tau^a\zeta({\bf z})\rangle,
  &\enum\cr
  \noalign{\vskip2ex}
  \fp(x_0)&=-a^6\sum_{\bf y,z}
  \frac{1}{3}\langle P^a(x)\,
  \zetabar({\bf y})\dirac{5}\frac{1}{2}\tau^a\zeta({\bf z})\rangle.
  &\enum\cr}
$$
Here and in the following we adopt the convention that
repeated indices are summed over.
To study the improved vector current~(2.5) we also introduce
the new correlation functions,
$$
  \eqalignno{
  \kv(x_0)&=-a^6\sum_{\bf y,z}
  \frac{1}{9}\langle V_k^a(x)\,
  \zetabar({\bf y})\dirac{k}\frac{1}{2}\tau^a\zeta({\bf z})\rangle,
  &\enum\cr
  \noalign{\vskip2ex}
  \kt(x_0)&=-a^6\sum_{\bf y,z}
  \frac{1}{9}\langle T_{k0}^a(x)\,
  \zetabar({\bf y})\dirac{k}\frac{1}{2}\tau^a\zeta({\bf z})\rangle.
  &\enum\cr}
$$
The amplitudes above are sufficient for the
determination of most of the
improvement coefficients of sect.~2.
However, in order to carry out a few additional 
checks we also calculated the boundary-to-boundary 
correlation~$\f1$,
$$
  \f1 =-{{a^{12}}\over{L^6}}
   \sum_{\bf u,v,y,z}
   \frac{1}{3}
   \langle\zetabarprime({\bf u})
   \dirac{5}\frac{1}{2}\tau^a\zetaprime({\bf v})
   \zetabar({\bf y})\dirac{5}\frac{1}{2}\tau^a\zeta({\bf z})\rangle,
  \eqno\enum
$$
which has previously appeared in the normalization conditions for
the isovector axial vector and vector currents in 
refs.~[\ref{letter},\ref{paperIV}].
As an aside we recall that $\f1$ will
also be needed for the  computation of the 
running quark mass in the $\SF$ scheme as defined in ref.~[\ref{letter}].
A detailed discussion of this topic will be presented 
elsewhere~[\ref{StefanPeter},\ref{MartinRainerHartmut}].

\subsection 3.2 Integration over the quark fields

The fermionic action being bilinear in the quark fields, the
corresponding Grassmann integration can be carried out
analytically using Wick's theorem.
To write down the resulting expressions in a compact form
we follow ref.~[\ref{paperII}] and introduce the matrix $H(x)$
through
$$
  H(x)=a^3\sum_{\bf y}
       {\delta\psiclass(x)\over\delta\rho({\bf y})}.
  \eqno\enum
$$
Here $\psiclass$ denotes the classical solution 
of the Dirac equation in a given 
gauge field configuration and $\rho$ is its boundary value
at time $x_0=0$.
One then finds
$$
  \eqalignno{
  \fa(x_0) &=-\frac{1}{2}\left\langle\tr\!\left\{
     H(x)^{\dagger}\dirac{0}H(x)\right\}\right\rangle_{\rm G},&\enum\cr
  \noalign{\vskip2ex}
  \fp(x_0) &=\frac{1}{2}\left\langle\tr\!\left\{
     H(x)^{\dagger}H(x)\right\}\right\rangle_{\rm G},&\enum\cr
  \noalign{\vskip2ex}
  \kv(x_0) &=-\frac{1}{6}\left\langle\tr\!\left\{\dirac{5}\dirac{k}
     H(x)^{\dagger}\dirac{5}\dirac{k} H(x)\right\}\right
     \rangle_{\rm G},&\enum\cr
  \noalign{\vskip2ex}
  \kt(x_0) &=\frac{1}{6}\left\langle\tr\!\left\{\dirac{5}\dirac{k}
     H(x)^{\dagger}\dirac{5}\dirac{k}\dirac{0} H(x)\right\}\right
     \rangle_{\rm G},
  &\enum\cr}
$$
where the trace is over the Dirac and colour indices. 
The bracket  $\langle\ldots\rangle_{\rm G}$ means that 
expectation values have to be taken with the effective gauge field measure  
including the fermionic determinant.

In order to obtain a compact expression for $\f1$
we follow ref.~[\ref{paperIV}] and introduce the matrix 
$$
  K = \ctildet{{a^3}\over{L^3}}\sum_{\bf x}
      \Bigl\{P_+U(x,0)^{-1}H(x)\Bigr\}_{x_0=T-a}.
  \eqno\enum
$$
Then one has
$$
  \f1=\frac{1}{2}\left\langle\tr\!\left\{K^{\dagger}K\right\}\right
     \rangle_{\rm G}.
  \eqno\enum
$$

\subsection 3.3 Perturbation expansion

The perturbation expansion is now easily generated following
ref.~[\ref{paperII}]. In particular we choose
vanishing boundary gauge fields $C$ and $C'$ and take over
the corresponding gauge fixing procedure.
The gluon field is introduced in the standard way 
by parameterizing  the link variables according to
$$
  U(x,\mu)=\exp\{ag_0 q_\mu(x)\}.
  \eqno\enum
$$
As explained in ref.~[\ref{paperII}], the classical
quark field and thus $H(x)$ can be expanded in perturbation theory, 
$$
  H(x)=H^{(0)}(x)+g_0 H^{(1)}(x)+g_0^2 H^{(2)}(x) + \rmO(g_0^3).
  \eqno\enum
$$
Also expanding the boundary improvement coefficient~[\ref{paperI}],
$$
  \ctildet=1+\ctildet^{(1)}g_0^2+\rmO(g_0^4),
  \eqno\enum  
$$
the corresponding expansion for $K$ reads
$$
  K=K^{(0)}+g_0 K^{(1)}+g_0^2 K^{(2)} + \rmO(g_0^3),
  \eqno\enum
$$
with 
$$
 \eqalignno{
  K^{(0)} &={{a^3}\over{L^3}}\sum_{\bf x}
             P_+ H^{(0)}(T-a,{\bf x}), &\enum\cr
          \noalign{\vskip2ex}
  K^{(1)} &={{a^3}\over{L^3}}\sum_{\bf x}
             P_+\Bigl\{H^{(1)}(x)-aq_0(x)H^{(0)}(x)\Bigr\}_{x_0=T-a},
          &\enum\cr
          \noalign{\vskip2ex}
  K^{(2)} &={{a^3}\over{L^3}}\sum_{\bf x}
             P_+\Bigl\{H^{(2)}(x)-aq_0(x)H^{(1)}(x) \cr
              &\hphantom{012345678}
            +\frac{1}{2}[aq_0(x)]^2H^{(0)}(x)+\ctildet^{(1)}H^{(0)}(x) 
            \Bigr\}_{x_0=T-a}.
            &\enum\cr}
$$
Inserting these expansions in eqs.~(3.7)--(3.10),(3.12) and carrying
out the integrations over the gluon and ghost field variables
finally leads to the
desired expansion for the correlation functions,
$$
  f = f^{(0)}+g_0^2 f^{(1)} + \rmO(g_0^4),
  \eqno\enum
$$
where $f$ stands for any of the amplitudes 
in eqs.~(3.1)--(3.5).

\subsection 3.4 Tree level results

Explicit expressions for the tree level correlation functions $\fa^{(0)}$ 
and $\fp^{(0)}$ have been given in ref.~[\ref{paperII}]. 
The study of their approach 
to the continuum limit led to the determination of $\bzeta,\bm,\ba,\bp$
and $\ca$ at lowest order of perturbation theory.
To this order we have the relations 
$$
  \eqalignno{
  \kv^{(0)}(x_0)&=\frac23\fp^{(0)}(x_0)-\frac13\fa^{(0)}(x_0),&\enum\cr
  \noalign{\vskip2ex}
  \kt^{(0)}(x_0)&=\frac23\fa^{(0)}(x_0)-\frac13\fp^{(0)}(x_0),&\enum\cr
  \noalign{\vskip2ex}
  \f1^{(0)}&=\Bigl\{\frac12\fp^{(0)}(x_0)
                   -\frac12\fa^{(0)}(x_0)\Bigr\}_{x_0=T-a},
  &\enum\cr}
$$
and we then find the tree level results
$\bv^{(0)}=\bt^{(0)}=1$ and $\cv^{(0)}=0$. All coefficients
are thus known to lowest order of perturbation theory (cf.~sect.~2).
%
%

\subsection 3.5 Computation of $\kv^{(1)}$ and $\f1^{(1)}$ 

Except for the different Dirac structure 
the calculation of $\kv^{(1)}$ is completely analogous to the 
cases of $\fa^{(1)}$ and $\fp^{(1)}$ treated in ref.~[\ref{paperII}].
In particular, there are again three diagrams to be computed
which are the same as in fig.~1 of this reference
(where the cross in this case denotes the vector vertex).
The contribution of the quark boundary counterterm can also be
inferred from there by noting
$$
  \kv^{(1)}(x_0)_{\rm b}=
  \frac23\fp^{(1)}(x_0)_{\rm b}-\frac13\fa^{(1)}(x_0)_{\rm b}.
  \eqno\enum
$$
Note that the correlation function $\kt^{(1)}$ could be computed
along the same lines. However here, $\kt$ is only encountered
as an O($a$) counterterm for the correlation function of the
improved vector current~(2.5). Since the associated 
improvement coefficient $\cv$ vanishes at tree level 
we will only need the lowest order expression $\kt^{(0)}$ 
in the following. 

The correlation function $\f1$ differs from the 
cases treated previously and therefore deserves a more
detailed presentation. Inserting the expansion of
the matrix $K$~(3.16) in eq.~(3.12) we obtain 
$$
  \eqalignno{
  \f1^{(1)}&=
  \frac{1}{2}\bigl\langle\tr\!\bigl\{
  {K^{(1)}}^{\dagger}K^{(1)} 
  \bigr\}\bigr\rangle_{\tilde{\rm G}} 
  &\cr
  \noalign{\vskip2ex}
  &\kern0.4ex
  +\frac{1}{2}\bigl\langle\tr\!\bigl\{
   {K^{(2)}}^{\dagger}K^{(0)}  
  +{K^{(0)}}^{\dagger}K^{(2)}  
  \bigr\}\bigr\rangle_{\tilde{\rm G}}.
  &\enum\cr}
$$
Here, the bracket $\langle\ldots\rangle_{\tilde{\rm G}}$
means the integration over the gluon and ghost fields
as explained in ref.~[\ref{paperII}].
While the integration over the ghost fields is trivial at
this order of perturbation theory, the integration over
the gluon fields generates the diagrams displayed in
fig.~2.  The first three diagrams are similar to the ones computed 
for $\fa$,$\fp$ and $\kv$, and 
the explicit link variables in eq.~(3.11) lead to the 
additional diagrams~4~--~7.

\topinsert
\vbox{
\vskip0.0true cm

\centerline{
\epsfxsize=11 true cm
\epsfbox{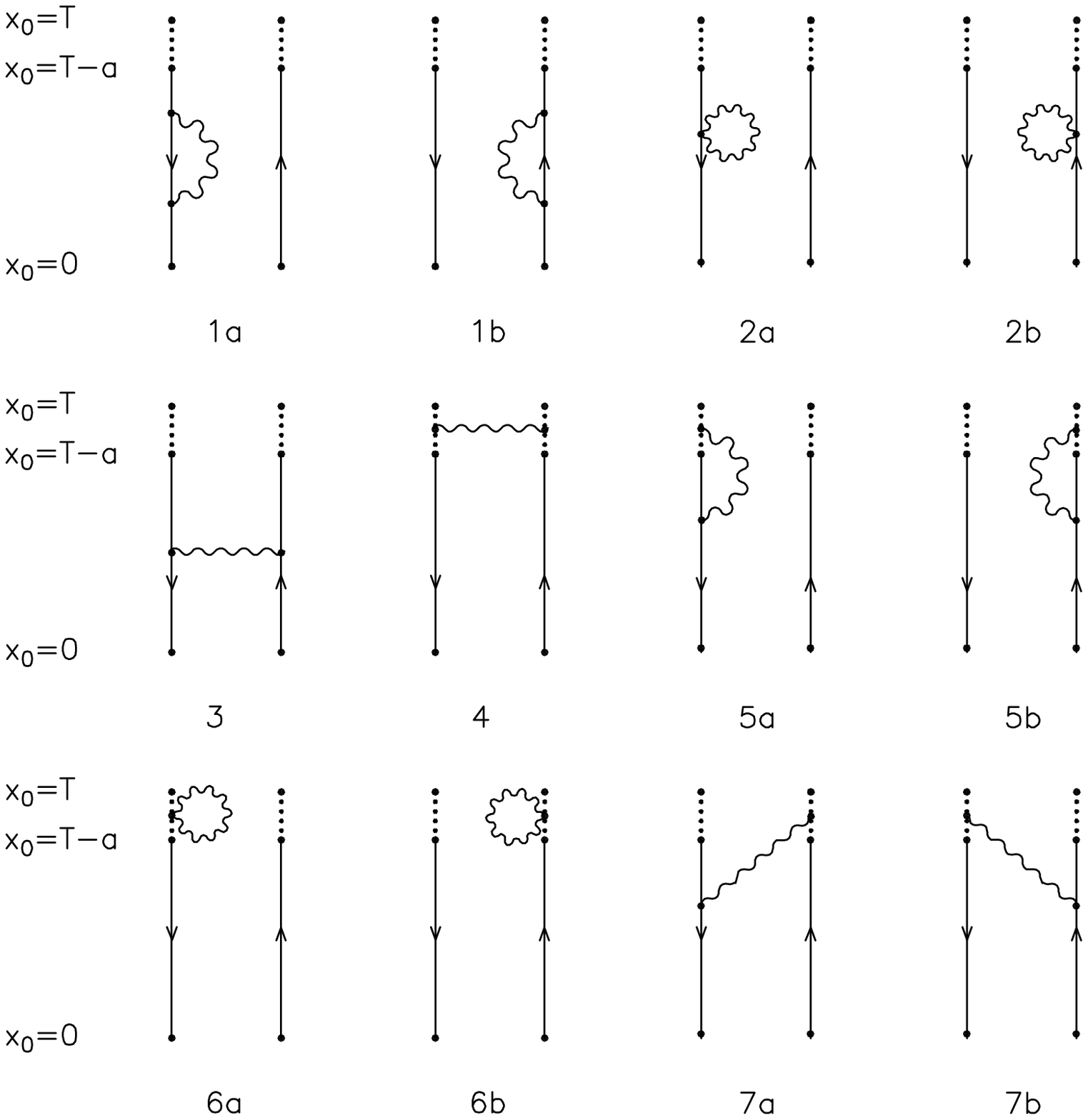}}

\vskip1.0true cm
\figurecaption{Diagrams contributing to $\f1$ at order $g_0^2$.
The dotted lines symbolize the link from Euclidean time $T-a$
to $T$.}
}
\endinsert

Furthermore, $\f1^{(1)}$ receives a contribution
from the quark boundary counterterm which can be written in
the form (see also eq.~(4.31) of ref.~[\ref{paperII}])
$$
 f_{1{\rm b}}^{(1)} = 2\ctildet^{(1)}\f1^{(0)}
 +\frac12 \fp^{(1)}(T-a)_{\rm b}-\frac12 \fa^{(1)}(T-a)_{\rm b}.
 \eqno\enum
$$
An explicit evaluation shows that this expression is indeed
of order $a$, as expected. 

On a lattice of a given size $T/a\times(L/a)^3$ the Feynman diagrams
can be evaluated numerically by inserting the
explicit time-momentum representation of 
the propagators and vertices
into the expressions for each diagram.
For this purpose two independent Fortran programs were written 
and the final results were checked against each other and 
for gauge invariance.
A fast version of one of the Fortran programs then enabled us to obtain
numerical results for a relatively 
large range of lattice sizes, thus allowing for 
rather precise extrapolations to the continuum limit (cf.~sect.~4).

\subsection 3.6 Renormalized correlation functions

In order to take the limit of large $L/a$ at fixed ratio $T/L$
and physical length $L$, the bare parameters 
and fields have to be scaled such that the renormalized
parameters and fields stay fixed. Furthermore this has to be done
in a way consistent with O($a$) improvement.
It has been shown in ref.~[\ref{paperI}] that 
these requirements are met by 
any mass-independent renormalization scheme with the 
properties as summarized in sect.~2. In particular,
the O($a$) improvement coefficients introduced there 
are then independent of the renormalization conditions.

In perturbation theory a 
particularly convenient choice is the minimal subtraction scheme
on the lattice which we shall adopt in the following.
In this scheme all renormalization constants at a given order in the
perturbation expansion in $\gtilde^2$ are polynomials 
in $\ln(a\mu)$ with mass-independent coefficients and no constant parts.

To first order of perturbation theory
the substitutions for the coupling constant and the quark mass 
then amount to
$$
  \eqalignno{
  g_0^2&=\gr^2+\rmO(\gr^4),
  &\enum\cr
 \noalign{\vskip2ex} 
  m_0&=m_0^{(0)}+\gr^2 m_0^{(1)}+\rmO(\gr^4),
  &\enum\cr}
$$
where the precise form of the coefficients,
$$
\eqalignno{
 m_0^{(0)}&={1\over a}\Bigl(1-\sqrt{1-2a\mr}\Bigr),
 &\enum\cr
 \noalign{\vskip2ex}
 m_0^{(1)}&=\mc^{(1)}
 -{{\zm^{(1)}\mr-2\bm^{(1)}\bigl(\mr-m_0^{(0)}\bigr)}
 \over{\sqrt{1-2a\mr}}},
 &\enum}
$$
is a direct consequence of the definitions made in sect.~2.

The renormalized correlation functions, defined by
$$
  \eqalignno{
  [\kv(x_0)]_{{\hbox{\sixrm R}}}&=
  \zv(1+\bv a\mq)\zzeta^2(1+\bzeta a\mq)^2
  &\cr
  \noalign{\vskip1ex}
  &\hphantom{\zv}\times
   \left\{\kv(x_0)+a\cv\frac12
  (\partial^*_0+\partial_0)\kt(x_0)\right\},
  &\enum\cr
  \noalign{\vskip2ex}
   [\fp(x_0)]_{{\hbox{\sixrm R}}}&=
  \zp(1+\bp a\mq)\zzeta^2(1+\bzeta a\mq)^2 \fp(x_0),
  &\enum\cr
  \noalign{\vskip2ex}
   [\fa(x_0)]_{{\hbox{\sixrm R}}}&=
  \za(1+\ba a\mq)\zzeta^2(1+\bzeta a\mq)^2
  &\cr
  \noalign{\vskip1ex}
  &\hphantom{\zv}\times
   \left\{\fa(x_0)+a\ca\frac12
  (\partial^*_0+\partial_0)\fp(x_0)\right\},
  &\enum\cr
  \noalign{\vskip2ex}
  [\f1]_{{\hbox{\sixrm R}}}&=
  \zzeta^4(1+\bzeta a\mq)^4 \f1,
  &\enum\cr}
$$
have a well-defined perturbation expansion in the renormalized
coupling $\gr$, with coefficients that are computable 
functions of $a/L$. However, to determine the
O($a$) improvement coefficients we only need to know
the renormalized amplitudes up to terms of order $a^2$.
Neglecting such terms, the expansion of $[\kv]_{{\hbox{\sixrm R}}}$ reads
$$
  \eqalignno{
  [\kv(x_0)]_{{\hbox{\sixrm R}}}&=
  \kv^{(0)}(x_0)+\gr^2\Bigl\{\kv^{(1)}(x_0)
  +m_0^{(1)}{\partial\over\partial m_0}\kv^{(0)}(x_0)
  &\cr
  \noalign{\vskip2ex}
  &\quad+\left(\zv^{(1)}+2\zzeta^{(1)}
    +a\mr\bigl[\bv^{(1)}+2\bzeta^{(1)}\bigr]\right)\kv^{(0)}(x_0)
  &\cr
  \noalign{\vskip2ex}
  &\qquad  +a\cv^{(1)}\frac12(\partial^*_0+\partial_0)
  \kt^{(0)}(x_0)\Bigr\}+\rmO(\gr^4),
  &\enum\cr}
$$
where it is understood that all quantities on the r.h.s.
are evaluated at $m_0=m_0^{(0)}$ as given in eq.~(3.29).
Analogous expressions are obtained for $[\fa]_{{\hbox{\sixrm R}}}$
and $[\fp]_{{\hbox{\sixrm R}}}$, in particular the case 
$\mr=0$ has already been discussed
in ref.~[\ref{paperII}]. 
We thus directly proceed to the result for $[\f1]_{{\hbox{\sixrm R}}}$,
$$
  \eqalignno{
  [\f1]_{{\hbox{\sixrm R}}}&=
  (1-2a\mr)\Bigl[\f1^{(0)}+\gr^2\Bigl\{\f1^{(1)}
  +m_0^{(1)}{\partial\over\partial m_0}\f1^{(0)}
  &\cr
  \noalign{\vskip2ex}
  &\quad+\left(4\zzeta^{(1)}
  +a\mr\bigl[4\bzeta^{(1)}+2\zm^{(1)}\bigr]\right)\f1^{(0)}\Bigr\}
  \Bigr]+\rmO(\gr^4).
  &\enum\cr}
$$
Since we are neglecting terms of order
$a^2$, the expansion 
$$
  m_0^{(1)}=\mc^{(1)}-\mr \Bigl[\zm^{(1)}+a\mr
  \Bigl(\zm^{(1)}+\bm^{(1)}\Bigr)\Bigr]+\rmO(a^2),
  \eqno\enum 
$$
may be inserted in eqs.~(3.35),(3.36).
However, note that 
we do not directly expand the coefficient $m_0^{(0)}$~[eq.~(3.29)]. 
One has to be careful here because the bare one-loop amplitudes
are linearly divergent. While the additive renormalization of
the quark mass removes these divergences, the correct evaluation of 
the renormalized amplitudes to O($a$) requires
a consistent treatment of $m_0^{(0)}$ to order $a^2$, 
in both the bare one-loop amplitudes and the mass counterterms.
It thus appears safer to first carry out the renormalization
procedure using the exact coefficient
and only neglect terms of O($a^2$) in the final 
result for the renormalized correlation functions.

\section 4. Computation of the improvement coefficients

The improvement coefficients can be determined
by requiring an improved continuum limit behaviour
of the renormalized correlation functions.
We first consider the case of vanishing renormalized quark
mass and compute $\cv^{(1)}$. We then turn to 
the massive case and also determine the $b$-coefficients.

\subsection 4.1 Choice of parameters

For the numerical analysis of the renormalized amplitudes
one needs to make a choice for the kinematical parameters
$T,L,x_0,\theta$ and the quark mass $\mr$.
Following ref.~[\ref{paperII}] we choose $T=2L$ and $x_0=T/2$.
Setting $z=\mr L$ we define the dimensionless functions,
$$
  \eqalignno{
  \hp(\theta,z,a/L)&=
  [\fp(x_0)]_{{\hbox{\sixrm R}}}\bigr|_{x_0=T/2},
  &\enum\cr
  \noalign{\vskip2ex}
  \ha(\theta,z,a/L)&=
  [\fa(x_0)]_{{\hbox{\sixrm R}}}\bigr|_{x_0=T/2},
  &\enum\cr
  \noalign{\vskip2ex}
  \hv(\theta,z,a/L)&=
  [\kv(x_0)]_{{\hbox{\sixrm R}}}\bigr|_{x_0=T/2},
  &\enum\cr
  \noalign{\vskip2ex}
  \hda(\theta,z,a/L)&=
  L\frac{1}{2}(\drvstar{0}+\drv{0})
  [\fa(x_0)]_{{\hbox{\sixrm R}}}\bigr|_{x_0=T/2},
  &\enum\cr
  \noalign{\vskip2ex}
  \h1(\theta,z,a/L)&=
  [\f1]_{{\hbox{\sixrm R}}}.
  &\enum\cr}
$$
From the discussion in sect.~3  one then infers the form of $\hp$, $\hda$
and $\h1$
$$
  \eqalignno{
  \hp&=u_0+\gr^2\biggl\{u_1+\ctildet^{(1)}u_2+a m_0^{(1)}u_3
  &\cr
  \noalign{\vskip2ex}
  &\quad
  +\left(\zp^{(1)}+2\zzeta^{(1)}+
    {{a}\over{L}}z\bigl[\bp^{(1)}+2\bzeta^{(1)}\bigr]\right)u_0
    \biggr\}+\rmO(\gr^4),
  &\enum\cr
  \noalign{\vskip4ex}
  \hda&=w_0+\gr^2\biggl\{w_1+\ctildet^{(1)}w_2+a m_0^{(1)}w_3+\ca^{(1)}w_4
  &\cr
  \noalign{\vskip2ex}
  &\quad
  +\left(\za^{(1)}+2\zzeta^{(1)}+
    {{a}\over{L}}z\bigl[\ba^{(1)}+2\bzeta^{(1)}\bigr]\right)w_0
    \biggr\}+\rmO(\gr^4),
  &\enum\cr
  \noalign{\vskip4ex}
   \h1&=t_0+\gr^2\biggl\{t_1+\ctildet^{(1)}t_2+a m_0^{(1)}t_3
  &\cr
  \noalign{\vskip2ex}
  &\quad +\left(4\zzeta^{(1)}
   +{{a}\over{L}}z\bigl[4\bzeta^{(1)}+2\zm^{(1)}\bigr]\right)t_0\biggr\}
   +\rmO(\gr^4).
  &\enum\cr}
$$
Similarly we find,
$$
  \eqalignno{
  \ha&=v_0+\gr^2\biggl\{v_1+\ctildet^{(1)}v_2+a m_0^{(1)}v_3+\ca^{(1)}v_4
  &\cr
  \noalign{\vskip2ex}
  &\quad
  +\left(\za^{(1)}+2\zzeta^{(1)}+
    {{a}\over{L}}z\bigl[\ba^{(1)}+2\bzeta^{(1)}\bigr]\right)v_0
    \biggr\}+\rmO(\gr^4),
  &\enum\cr}
$$
and the same for $\hv$, with the subscript A replaced by V and 
the coefficients $v_i$ replaced by $y_i$.

All the coefficients are still functions of $\theta$ and $z$.
Analytic expressions can be derived
for those coefficients which derive from the tree level
correlation functions or the O($a$) counterterms. 
Their expansions to order $a/L$ for the special case
of $z=0$ and non-vanishing $\theta$ are given in appendix~A
and in appendix~B of ref.~[\ref{paperII}].
In appendix B we give the corresponding results for
$\theta=0$ and non-vanishing renormalized quark mass.

The coefficients $v_1$,$y_1$,$u_1$,$w_1$ and $t_1$ are only obtained
numerically so that we must choose specific values for the 
parameters $\theta$ and $z$. 
In the case of vanishing quark mass $z=0$
we decided to collect numerical data for 
the three values $\theta=0$, $\theta=0.1$ and  $\theta=1$.
In the massive case we set $\theta=0$ 
and chose the three values $z=0.1$,
$z=0.5$ and $z=1$. 

With the above choices for the parameters the Feynman diagrams 
were then evaluated numerically in 128 bit precision arithmetic for
a sequence of lattice sizes ranging from $L/a=4$ to $L/a=48$.

\subsection 4.2 Logarithmic divergences

In the minimal subtraction scheme the one-loop renormalization constants
all take the form 
$$
  Z(\tilde{g}_0^2,a\mu)=1+Z^{(1)}\tilde{g}_0^2+\rmO(\tilde{g}_0^4),
  \eqno\enum
$$
where $Z^{(1)}$ is proportional to $\ln(a\mu)$.
From analytical results obtained with 
conventional perturbative techniques one then 
expects~[\ref{GabrielliEtAl}--\ref{StefanII},\ref{StefanI}]
$$
  \za^{(1)}=\zv^{(1)}=0,
  \eqno\enum
$$
and
$$
   \zp^{(1)}
  =-\zm^{(1)}
  =-2\zzeta^{(1)}={{6\CF}\over{(4\pi)^2}}\ln(a\mu).
  \eqno\enum
$$
In the course of our computation we were able to determine 
the coefficients of the logarithmic divergences 
numerically to high precision. The results were in complete
agreement with the above expectations and will not be
further discussed. In the numerical analysis presented below 
we set $\mu=1/L$ and use the exact coefficients when subtracting 
the logarithmically divergent parts.

\subsection 4.3 Computation of $\cv^{(1)}$

For vanishing quark mass the analogue of eq.~(4.9) for $\hv$ simplifies
to
$$
  \eqalignno{
  \hv(\theta,0,a/L)
  &=y_0+\gR^2\Bigl[ y_1+\ctildet^{(1)}y_2+a\mc^{(1)}y_3
  &\cr
  \noalign{\vskip2ex}
  &+\Bigl(\Zv^{(1)}+2\Zz^{(1)}\Bigr)y_0+\Cv^{(1)}y_4\Bigr]+\rmO(\gR^4).
  &\enum}
$$
In order to extract $\cv^{(1)}$
one can for example demand O($a$) improvement of the
combination $\hv-\frac13(2\hp-\ha)$. Due to the identities~(3.21)
and~(3.24) the coefficients $\ctildet^{(1)}$ and $a\mc^{(1)}$ 
do not contribute and the improvement condition 
then reads 
$$
  y_1-\frac13(2u_1-v_1)-\frac23\Zp^{(1)} u_0
  -\frac13(\Cv^{(1)}-\Ca^{(1)})v_4
  ={\rm const}+\rmO(a^2).
  \eqno\enum
$$
Here we have anticipated the absence of O($a$) effects 
proportional to $\ln(L/a)$, which are  
cancelled by setting the coefficient of the 
Sheikholeslami-Wohlert term to $\csw=1$. We have 
checked in a number of cases that this indeed happens.
Defining
$$
  {\cal C}(L)=
  y_1-\frac13(2u_1-v_1)
  +{{\cf}\over{4\pi^2}}u_0\ln(L/a),
  \eqno\enum
$$
and the symmetric difference operator
$$
  \frac12(\partial+\partial^*){\cal C}(L)
  ={{1}\over{2a}}\Bigl[{\cal C}(L+a)-{\cal C}(L-a)\Bigr],
  \eqno\enum
$$
we find
$$
  \cv^{(1)}-\ca^{(1)}=
  -\lim_{L/a\to\infty} {{3L}\over{v_4}}
  \frac12(\partial+\partial^*){\cal C}(L).
  \eqno\enum
$$
It turns out that the $\rmO(a/L)$ corrections in the expression above
are large for both values of $\theta$ considered and the 
extrapolation must therefore be done carefully. 
We used the method described in ref.~[\ref{LueWe}], appropriately
adapted to the present case and finally obtained 
$$
  \cv^{(1)}-\ca^{(1)}=-0.00657(1)\times\cf.
  \eqno\enum
$$
Insertion of the numerical value for $\ca^{(1)}$ [eq.~(2.15)]
then leads to the result quoted in eq.~(2.16).

An alternative determination of $\cv^{(1)}$ can be obtained by
requiring O($a$) improvement 
of the ratio $\hv(\theta,0,a/L)/\hv(0,0,a/L)$.
With the results of appendix~A this is equivalent
to demanding
$$
  y_1-{1\over N}y_0(y_1)_{\theta=0}^{}
  +a\mc^{(1)}\Bigl(y_3+2y_0{L\over a}\Bigr)
  +\cv^{(1)}y_4+\ctildet^{(1)}y_2
  ={\rm const}+\rmO(a^2).
  \eqno\enum
$$
While the renormalization constants cancel in this ratio we
now need to know the one-loop coefficient $\mc^{(1)}$
of the critical bare quark mass. We here 
follow ref.~[\ref{paperII}] and use the relation
$$
  (w_1/w_3)_{\theta=z=0}^{} = -a\mc^{(1)}+\rmO(a^3/L^3).
  \eqno\enum
$$
We then define the combination
$$
  {\cal C}'(L)=y_1-{1\over N}y_0(y_1)_{\theta=0}
  -\Bigl(y_3+2y_0{L\over a}\Bigr)(w_1/w_3)_{\theta=z=0}^{},
  \eqno\enum
$$
where it is understood that $w_1$ and $w_3$ are computed 
at the same lattice size as the other coefficients.
With the results of appendix~A we then find
$$
  \cv^{(1)}+4{{\cosh(2\sqrt{3}\,\theta)+1}\over{\cosh(2\sqrt{3}\,\theta)}}
  \ctildet^{(1)}=
  \lim_{L/a\to\infty} {{L}\over{y_4}}
  \frac12(\partial+\partial^*){\cal C}'(L) .
  \eqno\enum
$$
Evaluation of this relation at two different values of $\theta$ 
allows to eliminate $\ctildet^{(1)}$ and thus
leads to a direct determination of $\cv^{(1)}$.
Complete consistency with the previous method was found. 

With our extended data we now 
also obtain a slightly more accurate estimate
of $\ctildet^{(1)}$ than the one given in ref.~[\ref{paperII}], viz.
$$
  \ctildet^{(1)} = -0.01346(1)\times\cf.  
  \eqno\enum
$$

\subsection 4.4 Computation of the $b$-coefficients to one-loop order

In order  to compute the $b$-coefficients to one-loop order we now
consider the massive case where $z=\mr L$ is kept fixed at a non-zero 
value. In the following we always assume $\theta=0$  and use
the coefficients given in appendix~B.

We start by considering $\h1$~[eq.~(4.8)], 
which can be written in the form
$$
  \h1=N\rme^{\displaystyle-4z}\Bigl[1+\gr^2\Bigl\{{\cal T}(L)
  +4z\bigl(\bzeta^{(1)}+\ctildet^{(1)}+z\bm^{(1)}\bigr){a\over{L}}
  \Bigr\}+\rmO(\gr^4)\Bigr].
  \eqno\enum
$$
Using the exact results for the logarithmic divergences 
we have defined
$$
  {\cal T}(L)=N^{-1}\rme^{\displaystyle 4z}\Bigl[ t_1
  +a\mc^{(1)}t_3\Bigr]
  +2(1+2z){{6\cf}\over{(4\pi)^2}}\ln (L/a).
  \eqno\enum
$$
To evaluate ${\cal T}(L)$ 
numerically on a lattice of size $L/a$,
we here again follow the procedure above and replace 
the coefficient $a\mc^{(1)}$ by the ratio $-(w_1/w_3)_{\theta=z=0}$,
computed for the same lattice size. 
Requiring improvement of $\h1$
then determines the linear combination of improvement coefficients,
$$
  4z\Bigl[ \ctildet^{(1)}+\bzeta^{(1)}+z\bm^{(1)}\Bigr]
  =\lim_{L/a\to\infty} (L^2/a)\frac12(\partial+\partial^*) {\cal T}(L).
  \eqno\enum
$$
Since $\ctildet^{(1)}$ is already known, $\bzeta^{(1)}$
and $\bm^{(1)}$ can be determined by evaluating the 
improvement condition at two different values of the quark mass.

We can proceed similarly for the other correlation functions
$\fp,\fa$ and $\kv$. We here require the improved continuum
limit behaviour for the ratios $\hp^2/\h1$, $\ha^2/\h1$, 
and $\hv^2/\h1$.
Defining
$$
\eqalignno{
  {\cal U}(L)&= \phantom{-}N^{-1}\rme^{\displaystyle2z}
  \Bigl[u_1+a\mc^{(1)}u_3\Bigr]
  +2z{{6\cf}\over{(4\pi)^2}}\ln (L/a),
  &\enum\cr
  \noalign{\vskip2ex}
  {\cal V}(L)&= -N^{-1}\rme^{\displaystyle 2z}\Bigl[v_1+a\mc^{(1)}v_3\Bigr]
  +(1+2z){{6\cf}\over{(4\pi)^2}}\ln (L/a),
  &\enum\cr
  \noalign{\vskip2ex}
  {\cal Y}(L)&= \phantom{-}N^{-1}\rme^{\displaystyle 2z}
  \Bigl[y_1+a\mc^{(1)}y_3\Bigr]
  +(1+2z){{6\cf}\over{(4\pi)^2}}\ln (L/a),
  &\enum}
$$
we find e.g.
$$
  {1\over N}{\hv^2\over\h1} = 1+\gr^2\Bigl[2{\cal Y}(L)-{\cal T}(L)
  +2z{a\over L}\bigl(2\cv^{(1)}+\bv^{(1)}\bigr)\Bigr]+\rmO(\gr^4).
  \eqno\enum
$$
Similar expressions can be derived for the other cases and we
thus find that O($a$) improvement requires
$$
\eqalignno{
  z\bp^{(1)}
  &=\lim_{L/a\to\infty} (L^2/a)\frac12(\partial+\partial^*)
  \Bigl[ {\cal U}(L)-\frac12 {\cal T}(L)\Bigr],
  &\enum\cr
  \noalign{\vskip2ex}
  z\Bigl[2\Ca^{(1)}+\ba^{(1)}\Bigr]
  &=\lim_{L/a\to\infty} (L^2/a)\frac12(\partial+\partial^*)
  \Bigl[ {\cal V}(L)-\frac12 {\cal T}(L)\Bigr],
  &\enum\cr
  \noalign{\vskip2ex}
  z\Bigl[2\Cv^{(1)}+\bv^{(1)}\Bigr] 
  &=\lim_{L/a\to\infty} (L^2/a)\frac12(\partial+\partial^*)
  \Bigl[ {\cal Y}(L)-\frac12 {\cal T}(L)\Bigr].
  &\enum}
$$
The evaluation of the various improvement conditions 
was done using the methods of ref.~[\ref{LueWe}] and
led to the numerical results as quoted in table~1.
Various consistency checks were made by studying different
combinations of correlation functions.
For example, a more direct determination 
of $\bm^{(1)}$ was obtained by
considering the ratio $\hda(0,z,a/L)/\ha(0,z,a/L)$.
As a further confirmation of the conceptual framework, 
we have also analyzed data at finite renormalized quark mass 
and values of $\theta\ne0$ and verified 
that the $b$-coefficients are indeed independent of $\theta$.

\section 5. Concluding remarks

This paper presents a further step in a systematic
investigation of the continuum limit of lattice QCD.
At one-loop order of perturbation theory
we have studied the continuum limit of 
many different renormalized amplitudes.
Various non-trivial consistency checks
have been carried out, for both zero
and non-zero renormalized quark mass.
As a result our analysis further supports the consistency of the
general framework of on-shell O($a$) improvement as presented
in ref.~[\ref{paperI}].  

The O($a$) improvement 
coefficients which enter the definition of
the improved lattice action and the quark bilinear isovector
pseudo-scalar, vector and axial vector operators are now
all known to one-loop order of perturbation theory.
The ultimate aim is of course to compute the same coefficients
non-perturbatively. This has been partially achieved
for the quenched theory (i.e.~for $\Nf=0$) in 
refs.~[\ref{paperIII},\ref{paperIV}]. More precisely,
the coefficients $\csw,\ca$ and $\bv$ have been 
determined for this case and a similar determination of 
$\cv$ is under way~[\ref{MarcoRainer}]. 
The extension to the full theory
with two quark flavours has also been initiated
and first results are expected in the near future~[\ref{KarlRainer}].

However, the non-perturbative 
methods developed so far cannot be used in a straightforward
manner to determine all of the improvement coefficients. 
An example is the coefficient $\ba$ for which
one presently has to rely on perturbative estimates. 
Of course, this does not represent
a severe limitation provided the quark mass 
(measured in lattice units) is small enough. In this context 
it may be of some interest  that the perturbative 
results for $\bv$ and $\ba$ are approximately of the same magnitude. 
While this need not remain true at higher orders 
or beyond perturbation theory it is certainly tempting 
to use the non-perturbative result for $\bv$ as 
a first guess for $\ba$.

\subsection Acknowledgements

This work is part of the ALPHA collaboration research programme.
We would like to thank our colleagues Martin L\"{u}scher,
Rainer Sommer and Hartmut Wittig for helpful discussions
and critical comments on a first draft of this paper.
Stefan Sint thanks the Max-Planck-Institut 
for hospitality and acknowledges support by the U.S.~Department of
Energy (contracts DE-FG05-85ER250000 and DE-FG05-96ER40979).
\vfill
\eject

\appendix A

For the massless case $z=\mr L=0$, the 
expansions of the functions $u_i$,$v_i$ and $w_i$
(for $i\ne 1$) have been given in appendix B of ref.~[\ref{paperII}],
up to corrections of order $(a/L)^2$.
Our definitions then imply
$$
  \eqalignno{
  y_i&=\frac23 u_i-\frac13 v_i \qquad {\rm for}\quad i=0,2,3,
  &\enum\cr
  \noalign{\vskip2ex}
  y_4&=-\frac13 v_4 +\frac23 w_0{a\over L}.
  &\enum}
$$
Using the abbreviations
$$
  {\rm co}=\cosh(2\sqrt{3}\,\theta),\qquad 
  {\rm si}=\sinh(2\sqrt{3}\,\theta),
  \eqno\enum
$$
we also give the explicit expressions,
$$
\eqalignno{
  y_0&={N(2\co+1)\over 3\co^2},
  &\enum\cr
  \noalign{\vskip2ex}
  y_2&={8N\theta\si(\co+1)\over \sqrt{3}\co^3}{a\over L},
  &\enum\cr
  \noalign{\vskip2ex}
  y_3&=-{N\si(\co+2)\over3\sqrt{3}\theta\co^3}{L\over a}
     -\Bigl\{ {N\theta\si(\co+2)\over 18\sqrt{3}\co^3}
  &\cr
  \noalign{\vskip2ex}
  &+{4N\theta^2\over 9\co^4}(\co^3+4\co^2-2\co-6)\Bigr\}{a\over L},
  &\enum\cr
  \noalign{\vskip2ex}
  y_4&={2N\theta\si\over\sqrt{3}\co^2}{a\over L}.
  &\enum}
$$
For the functions $t_i$ introduced in eq.~(4.8)
we obtain
$$
\eqalignno{
  t_0&={N\over\co^2},
  &\enum\cr
  \noalign{\vskip2ex}
  t_2&={24N\theta\si\over \sqrt{3}\co^3}{a\over L},
  &\enum\cr
  \noalign{\vskip2ex}
  t_3&=-{2N\si\over\sqrt{3}\theta\co^3}{L\over a}
  +{2N\over\co^2}
  &\cr
  \noalign{\vskip2ex}
  &-\Bigl\{ {19N\theta\si\over3\sqrt{3}\co^3}
  -{8N\theta^2(1-2\si^2)\over3\co^4}\Bigr\}{a\over L}.
  &\enum}
$$

\appendix B

In this appendix we provide explicit expressions
for the coefficient functions introduced in eqs.~(4.6)--(4.9), for 
the special case $\theta=0$. 
We set $z=\mr L$ and use the tree level expression of 
the renormalized quark mass, $\mr=m_0(1-am_0/2)$.
Up to terms of order $a^2/L^2$ we then find
$$
  \eqalignno{
  t_0&=N\rme^{\displaystyle -4z}\,,
  &\enum\cr
  \noalign{\vskip2ex}  
  t_2&=N\rme^{\displaystyle -4z}\,4z{a\over L},
  &\enum\cr
  \noalign{\vskip2ex}
  t_3&=N\rme^{\displaystyle -4z}\,\Bigl\{ -4{L\over a}+2+4z
  +2z\bigl(\frac83z^2 +3z-1\bigr){a\over L}\Bigr\},
  &\enum\cr
  \noalign{\vskip3ex}
  u_0&=N\rme^{\displaystyle -2z}\,,
  &\enum\cr
  \noalign{\vskip2ex}
  u_2&=N\rme^{\displaystyle -2z}\,2z{a\over L},
  &\enum\cr
  \noalign{\vskip2ex}
  u_3&=N\rme^{\displaystyle -2z}\,\Bigl\{-2{L\over a}+2z
  +z^2\bigl(\frac43z-1\bigr){a\over L}\Bigr\},
  &\enum\cr
  \noalign{\vskip3ex}
  v_i&=-u_i\quad\quad {\rm for}\quad\quad i=0,2,3,
  &\enum\cr
  \noalign{\vskip2ex}
  v_4&=-N\rme^{\displaystyle -2z}\,2z{a\over L},
  &\enum\cr
  \noalign{\vskip2ex}
  w_0&=N\rme^{\displaystyle -2z}\,2z,
  &\enum\cr
  \noalign{\vskip2ex}
  w_2&=N\rme^{\displaystyle -2z}\,4z^2{a\over L},
  &\enum\cr
  \noalign{\vskip2ex}
  w_3&=N\rme^{\displaystyle -2z}\,\Bigl\{\bigl(2-4z\bigr){L\over a}
  +4z^2-2z
  &\cr
  &\hphantom{01234567}
    +z^2\bigl(5-\frac{22}3z+\frac83z^2\bigl){a\over L}\Bigr\},
  &\enum\cr
  \noalign{\vskip2ex}
  w_4&=N\rme^{\displaystyle -2z}\,4z^2{a\over L},
  &\enum\cr
  \noalign{\vskip3ex}
  y_i&=-v_i\quad\quad {\rm for}\quad\quad i=0,2,3,4.
  &\enum}
$$


\beginbibliography

\bibitem{letter}
K. Jansen, C. Liu, M. L\"{u}scher, H. Simma, S. Sint,
R. Sommer, P. Weisz and U. Wolff,
Phys. Letts. B372 (1996) 275

\bibitem{latconf}
M. L\"{u}scher, S. Sint, R. Sommer, P. Weisz, H. Wittig
and U. Wolff, Nucl. Phys. B (Proc. Suppl.) 53 (1997) 905
 
\bibitem{paperI}
M. L\"uscher, S. Sint, R. Sommer and P. Weisz,
Nucl. Phys. B478 (1996) 365

\bibitem{paperII}
M. L\"uscher and P. Weisz,
Nucl. Phys. B479 (1996) 429

\bibitem{paperIII}
M. L\"uscher, S. Sint, R. Sommer, P. Weisz and U. Wolff,
Non-perturbative O($a$) improvement of lattice QCD,
hep-lat/9609035, accepted for publication in Nucl. Phys. B

\bibitem{paperIV}
M. L\"uscher, S. Sint, R. Sommer and H. Wittig,
Non-perturbative determination of the axial current normalization constant
in O($a$) improved lattice QCD, 
hep-lat/9611015, accepted for publication in Nucl. Phys. B

\bibitem{alphaI}
M. L\"uscher, R. Narayanan, P. Weisz and U. Wolff,
Nucl. Phys. B384 (1992) 168

\bibitem{StefanI}
S. Sint, 
Nucl. Phys. B421 (1994) 135,
Nucl. Phys. B451 (1995) 416

\bibitem{SW}
B. Sheikholeslami and R. Wohlert,
Nucl. Phys. B259 (1985) 572

\bibitem{Wohlert}
R. Wohlert, Improved continuum limit lattice action for quarks,
DESY preprint 87-069 (1987), unpublished

\bibitem{StefanRainer}
S. Sint and R. Sommer,
Nucl. Phys. B465 (1996) 71

\bibitem{Parisi}
G. Parisi, in: High--Energy Physics --- 1980, XX. Int. Conf.
Madison (1980), ed. L. Durand and L. G. Pondrom
(American Institute of Physics, New York, 1981)

\bibitem{StefanPeter}
S. Sint and P. Weisz, work in progress

\bibitem{MartinRainerHartmut}
M. L\"{u}scher, R. Sommer and H. Wittig, work in progress


\bibitem{GabrielliEtAl}
E. Gabrielli, G. Martinelli, C. Pittori, G. Heatlie and C. T. Sachrajda,
Nucl. Phys. B362 (1991) 475

\bibitem{GoeckelerEtAl}
M. G\"{o}ckeler et al., Nucl. Phys. B (Proc. Suppl.) 53 (1997) 896

\bibitem{StefanII}
S. Sint, private notes (1993,1996)


\bibitem{LueWe}      
M. L\"uscher and P. Weisz, 
Nucl. Phys. B266 (1986) 309

\bibitem{MarcoRainer}
M. Guagnelli and R. Sommer, work in progress

\bibitem{KarlRainer}
K. Jansen and R. Sommer, work in progress

\endbibliography

\bye